\documentclass[runningheads]{svmult}

\usepackage{graphicx}  
\usepackage{subeqnar}  
\usepackage{cropmark} 
\usepackage{physprbb}  

\newcommand{\beq}{\begin{equation}}
\newcommand{\eeq}{\end{equation}}
\newcommand{\bea}{\begin{eqnarray}}
\newcommand{\eea}{\end{eqnarray}}
\begin{document}
\title*{ 
     Baryon stopping in high energy collisions in the DPMJET--III model
}
\toctitle{
     Baryon stopping in high energy collisions in the DPMJET--III model
}
%
%
\titlerunning{
     Baryon stopping in high energy collisions 
}
%
 \author{  J.~Ranft\inst{1} 
\and R.~Engel\inst{2}
\and  S.~Roesler\inst{3}} 
\authorrunning{J.~Ranft et al.}
%
%
\institute{ Physics Dept. Universit\"at Siegen, D--57068 Siegen, Germany,
 e--mail: Johannes.Ranft@cern.ch
\and University of Delaware, Bartol Res. Inst., Newark DE 19716 USA,
 e--mail: eng@lepton.bartol.udel.edu
\and SLAC, P.O. Box 4349, Stanford CA 94309 USA,
 sroesler@slac.stanford.edu}

\maketitle              
 \vspace{4mm}
{\bf SLAC--PUB--8734}\\
{\em To be published in the proceedings of the meeting Monte Carlo 2000,
Lisboa}

 \vspace{4mm}

 A recently discovered  feature of hadron production in  nuclear
 collisions is the large stopping of the
 participating nucleons 
 in hadron--nucleus and nucleus--nucleus collisions. Experimental data
 demonstrating this effect have been presented in
 \cite{NA35FIN,Alber98}. 

Multistring fragmentation models like the Dual Parton Model (DPM)
or similar models
did originally not show this enhanced stopping in nuclear collisions.
Therefore, in order to incorporate the effect into multistring
fragmentation models
 new diquark breaking DPM--diagrams acting in
 hadron--nucleus and nucleus--nucleus collisions were 
 proposed by  Kharzeev \cite{Kharzeev96} and Capella and Kopeliovich
 \cite{Capella96} and investigated in detail by Capella and
 collaborators \cite{Capella99a,Capella99}. 
 Similar ideas were discussed by Vance and
 Gyulassy\cite{Vance99}
   and by Casado \cite{Casado99}.
 
The  Monte Carlo implementation into DPMJET--II.5 of the new
diquark breaking diagrams of   Kharzeev \cite{Kharzeev96} and Capella
and Kopeliovich  \cite{Capella96} was first discussed by Ranft
\cite{Ranft20001} . 
The implementation into  DPMJET--III \cite{Roesler20001}
of these diagrams 
 differs somewhat from
\cite{Ranft20001} and is described here.

There are 
 two possibilities for the first fragmentation step of a diquark.
 Either we
get in the first step a baryon, which contains
both quarks of the diquark and the string junction or   in the
first step a meson is produced containing only one of the two quarks and the
baryon is produced in one of  
the following fragmentation steps.
 This mechanism was implemented
 under the name {\it popcorn}
fragmentation in the Lund chain fragmentation model JETSET
\cite{JETSET,AND85} which is presently used in DPMJET. 
The popcorn mechanism  alone is  not enough to explain the baryon
stopping observed experimentally in
hadron--nucleus and nucleus--nucleus
collisions
 \cite{NA35FIN,Alber98}.

In Ref.\cite{Ranft20001} we describe these new  diquark breaking
DPM--diagrams  in detail. This
will not be repeated here. The two important diagrams are 

(i)GSQBS, the Glauber sea quark mechanism of baryon stopping, this
diagram acts in nuclear collisions already at low energy.

(ii)USQBS, the unitary sea quark mechanism of baryon stopping, this
mechanism leads to baryon stopping also in proton--proton collisions at
collider and cosmic ray energies.

\begin{figure}[thb] \centering
\begin{center}
\includegraphics[width=12.0cm,height=3.0cm]{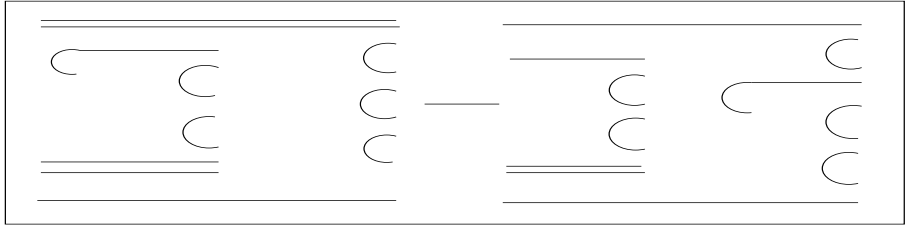}
\end{center}
 \caption[]{Diquark breaking in an original diquark--quark and
 seaquark--diquark chain system.}
 \label{dia1}
 \end{figure}

In DPMJET--III we first construct the system of parton chains according
to the model without the diquark breaking diagrams. Having this we
search for situations (i) as plotted on the left hand side of Fig.
\ref{dia1} or situations (ii) (not plotted),
 where the left lower diquark is replaced by
an antiquark.
On the left hand side in Fig. \ref{dia1} we have a 
diquark--quark chain and a
seaquark--diquark chain with the (upper) diquark and seaquark belonging
to the same projectile hadron and also the (lower) diquark and
valence--quark belonging to the same target--nucleon.
 In situation (ii) we have a diquark--quark chain and a
seaquark--anti--seaquark chain with the (upper) diquark 
and seaquark belonging
to the same projectile hadron and also the (lower) anti--seaquark and
valence--quark belonging to the same target--nucleon.
The  chain system is transformed as plotted 
on the right hand side of
Fig.\ref{dia1}. The projectile
diquark is split and the two resulting quarks 
come to the upper ends of both chains
and the projectile seaquark goes into the middle  of the second chain
and determines the position where the baryon is produced. The sea quarks
in Fig. \ref{dia1} might be Glauber sea quarks or
unitary sea quarks.

 Besides the situations already discussed  we
 consider also the ones where projectile and target are exchanged as
 well as situations with anti--diquarks.

We split the diquark by sampling for one of the two resulting valence
quarks (randomly choosen) a normal valence quark $x_{v1}$ and give to
the second valence quark $x_{v2} = x_d - x_{v1}$.
For each of the new diquark breaking diagrams (GSQBS and USQBS) 
we have to introduce a new parameter. These parameters give the
probability for the diquark breaking mechanisms to occur, 
given a suitable sea
quark is available and given that the diquark breaking mechanism is
kinematically allowed. For an original diquark--quark chain of small
invariant mass, which originally 
just fragments into two hadrons, the diquark
breaking is often not allowed.
The optimum values of the new parameters are determined by comparing
DPMJET--III with experimental data on net--baryon distributions. We 
obtain for the GSQBS  parameters the value 0.6 and use the same value for 
the USQBS parameter.

Introducing the new baryon stopping mechanisms into DPMJET we get an
 significant modification of the model in different sectors:
(i)The Feynman--$x$ distributions of leading protons in proton--proton
and proton--nucleus collisions.
 The leading particle
production is especially 
important for the cosmic ray cascade simulation.
(ii)The net--p ($p$ -- $\bar p$)  and net--$\Lambda$ ($\Lambda$ --
$\bar\Lambda$) rapidity distributions in hadron--nucleus and
nucleus--nucleus collisions. These are the data on the enhanced baryon
stopping mentioned already above.
(iii)The production of hyperons and anti--hyperons in nuclear
collisions.
We present here examples for (i) and (ii).

\begin{figure}[thb] \centering
\begin{center}
\includegraphics[width=8.0cm,height=4.0cm]{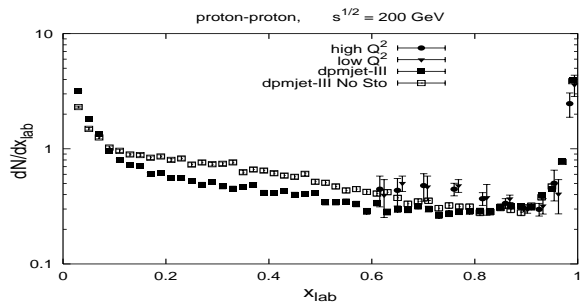}
\end{center}
 \caption{
Energy fraction \protect$x_{lab}$ carried by the leading proton.
The data are
photoproduction and DIS measurements at $\sqrt s \approx$  200 GeV
\protect\cite{Solano2000} 
 compared to  DPMJET--III with and without (No Sto) the 
diquark breaking diagrams for p--p
collisions at  $\sqrt s$ = 200 GeV. 
 \protect\label{xlab}
 }
 \end{figure}

 In Fig.  \ref{xlab} we compare 
the distribution in the energy fraction $x_{lab}$ 
carried by the leading proton. 
The data are
photoproduction and DIS measurements from the HERA collider 
at $\sqrt s \approx$  200 GeV
\cite{Solano2000}. We compare to DPMJET--III for p--p
collisions at  $\sqrt s $ = 200 GeV. The forward production of leading
protons is not expected to depend strongly on the reaction channel. 
We present the DPMJET--III distributions for the models 
with and without the
diquark breaking diagrams (No Sto in the plot).
The new diagrams modify the distributions mainly at intermediate
$x_{Lab}$ values where, unfortunately, no experimental data exist at
present. Therefore, further conclusions cannot be drawn at this point.

\begin{figure}[thb]
\begin{center}
\includegraphics[width=10.0cm,height=6.0cm]{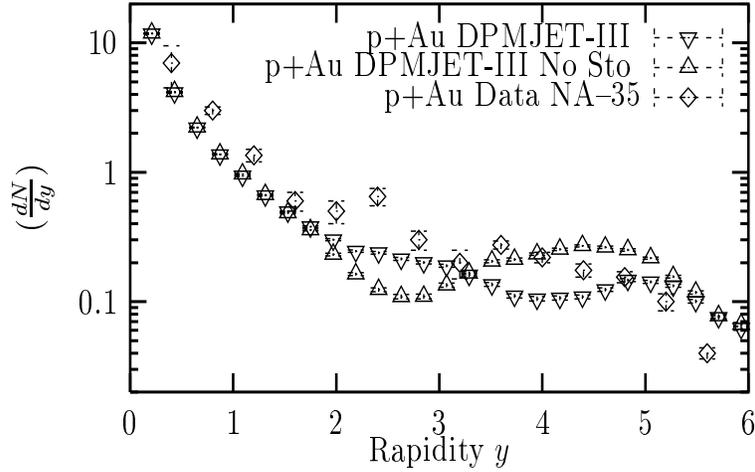}
\end{center}
\vspace*{-3mm}
\caption{
Net proton ($p-\bar p$) rapidity distribution in p--Au
collisions. 
The DPMJET--III results with and without (No Sto) 
the diquark breaking diagrams are compared with data
\protect\cite{Alber98}.
\protect\label{dpm3paunetp2}
}
\end{figure}

In Fig.  \ref{dpm3paunetp2} we compare the net--proton
distributions according to the   models with and without the diquark
breaking  diagrams with data in 
p--Au collisions \cite{Alber98}. 
The dip at central rapidity, which occurs in the model without the
baryon stopping diagrams  is filled.
The full model follows the relatively flat distribution at central
rapidities shown by the data.

\begin{figure}[thb]
\begin{center}
\includegraphics[width=10.0cm,height=6.0cm]{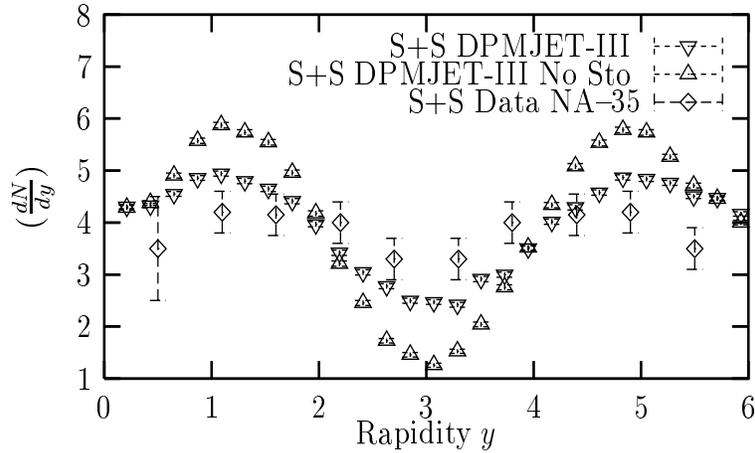}
\end{center}
\vspace*{-3mm}
\caption{
Net proton ($p-\bar p$) rapidity distribution in central 
S--S
collisions. 
The DPMJET--III results with and without (No Sto) 
the diquark breaking diagrams are compared with data
\protect\cite{Alber98}.
\protect\label{dpm3ssnetp2}
}
\end{figure}

In Fig. \ref{dpm3ssnetp2} we compare the  DPMJET--III model with and
without the diquark breaking diagrams 
with data
on net--proton production in central S--S collisions. 
 Also here the significant dip
at central rapidity in the model without the new diagrams 
is much less pronounced in the full model,
however, the agreement
to the data \cite{Alber98} is not perfect. 

The presence of the new 
baryon stopping diagrams modifies also 
the extrapolation of multistring models to cosmic ray energies. The
energy fractions carried by baryons decrease against those 
predicted by  models without the
new diagrams. The energy fractions carried by mesons and the spectrum
weighted moments of mesons increase as compared to
models without the new
diagrams. We present as function of the energy for two important
variables the predictions of DPMJET-III with and without the new baryon
stopping diagrams. All our plots are for p--p collisions, the model
behaves in a rather similar way also for p--Air  collisions.
We first discuss plots, where the baryon stopping mechanism causes
significant differences.

The cosmic ray spectrum--weighted moments  
are  defined as moments of $x_{lab}$ distribution for the production of
secondary particles $i$ in hadron--hadron and hadron--nucleus collisions

\begin{equation}
 F_i(x_{lab}) = x_{lab}\frac{dN_i}{dx_{lab}}
\end{equation}
	 
as follows

\begin{equation}
f_i = \int^{1}_0 (x_{lab})^{\gamma -1}
F_i(x_{lab})dx_{lab}.
\end{equation}
Here $-\gamma \simeq$ --1.7 is the power of the integral cosmic
ray energy spectrum.
The spectrum--weighted moments for nucleon--air collisions,
as discussed in ~\cite{gaistext}, determine the
uncorrelated fluxes of energetic particles in the atmosphere.

We also introduce the energy fraction $K_i$ :

\begin{equation}
K_i = \int^{1}_0 
F_i(x_{lab})dx_{lab}
\end{equation}
As for $x_{lab}$, the upper limit for $K$ is 1 in h--nucleus
collisions.

\begin{figure}[thb]
\begin{center}
\includegraphics[width=10.0cm,height=6.0cm]{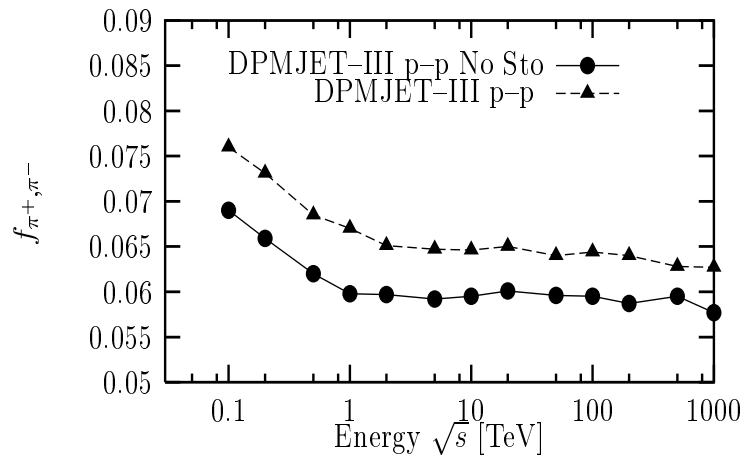}
\end{center}
\vspace*{-3mm}
\caption{Spectrum weighted moments for charged pion production in  p--p
 collisions as function of the 
cms energy  $\sqrt s$.
\protect\label{dpm3fpipp2}
}
\end{figure}

In Fig.\ref{dpm3fpipp2}  
we present the spectrum weighted moments for charged pions
 in  p--p collisions as function of
the cms energy $\sqrt s$.

\begin{figure}[thb]
\begin{center}
\includegraphics[width=10.0cm,height=6.0cm]{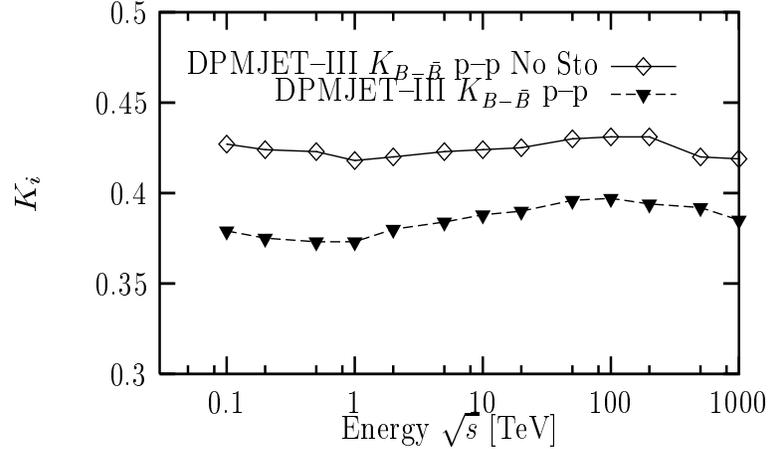}
\end{center}
\vspace*{-3mm}
\caption{
Laboratory energy fractions  for net baryons (baryon minus
antibaryon) $B-\bar B$
  production in p--p 
 collisions as function of the 
cms energy  $\sqrt s$.
\protect\label{dpm3kbpp2}
}
\end{figure}

In Fig.\ref{dpm3kbpp2}   
we present again for $pp$ 
collisions the energy fraction K  for net baryons $B-\bar B$ (baryon
minus
antibaryon). 
The
difference between  $K_{B-\bar B}$ and  $K_B$ 
is the energy fraction going into antibaryons 
$K_{\bar B}$  which is equal to the energy
fraction carried by the baryons which are newly 
produced in baryon--antibaryon
pairs.

There are also observables
 where the difference between the two versions
 of the model are rather insignificant. 
 Examples are the average transverse momentum of charged hadrons 
  as function of the energy and
 the average charged multiplicity $<n_{ch}>$ as
function of the collision energy.

{\bf Acknowledgements} The work of S.R and R.E. 
is supported  by the US Department of
Energy under contracts DE--AC03--76SF00515 and  DE--FG02 91ER40626,
respectively.

%

 \clearpage

%
 \end{document}